\documentclass[aip,amsmath,amssymb,reprint,groupedaddress,floatfix]{revtex4-1}
\usepackage[utf8]{inputenc}
\usepackage[T1]{fontenc}

\usepackage{graphicx}
\usepackage{floatrow}
\usepackage[caption=false,font={large}]{subfig}
\usepackage{bbm}

\usepackage{amsmath}
\usepackage{amssymb}
\usepackage{color}
\usepackage{times}
\usepackage{float}
\usepackage[sort&compress]{natbib}

\usepackage[colorinlistoftodos,prependcaption,textsize=small]{todonotes}

\usepackage{hyperref}

\hypersetup{
  colorlinks=true,
citecolor=black,
linkcolor=black,
urlcolor=black
  }

\newcommand{\egn}[1]{{#1}}

\begin{document}

\preprint{AIP/123-QED}

\date{\today}
\title{Stochastic models of memristive behavior}

\author{P. F. G\'ora}
\email{pawel.gora@uj.edu.pl}
 
\author{Ewa Gudowska-Nowak}

\affiliation{Institute of Theoretical Physics and Mark Kac Complex Systems Research Center, 
Jagiellonian University, \L{}ojasiewicza 11, 30-348 Krak\'ow, Poland}

\date{\today}

\begin{abstract}
Under normal operations, memristive devices undergo variability in time and space and have internal dynamics. Interplay of memory and stochastic signal processing in memristive devices makes them candidates for performing bio-inspired tasks of information transduction and transformation, where intrinsic random behavior can be harnessed for high performance of circuits built up of individual memory storing elements. The paper discusses models of single memristive devices exhibiting both - dynamic hysteresis and Stochastic Resonance, addressing also the cooperative effect of correlated noises acting on the system and occurrence of dirty hysteretic rounding.
\end{abstract}


\maketitle


{\bf The memristor, or the resistor with a memory, was first proposed by Chua in 1971 \cite{Chua1971} 
as a ``missing circuit element''.
The original idea has been further extended~\cite{Chua1976} to memristive systems (otherwise named resistance switching memory cells) and had gradually attracted 
interest from researchers and engineers, until the first actual memristor was reported to have been constructed
in 2008 in HP Labs \cite{Williams2008}. That incident brought about explosion of attention in the field of theoretical frameworks and fabrication techniques, both intended to devise and construct electronic memory structures.  There is a discussion whether true memristors, exhibiting the direct flux-charge interaction, exist. A discovery of such a device was reported in Ref.~\cite{Phi}, but this research was met with much criticism.~\cite{Pershin2022}, and has been later withdrawn. Usually memristive devices are described in terms of deterministic mathematical models in which a key element is 
the existence of a pinched hysteresis loop~\cite{Pershin2011,Chua_2014}. This feature of new components exhibiting memory storage captured special interest of the communities focused on information and communication technologies and found applications in areas such as unconventional, neuromorphic computing~\cite{Pershin_2019,Pershin2022}, machine learning, models of the brain
and many others; see Ref.~\cite{Caravelli_memristors_2018,Radwan2015,Battistoni,alonso_2021}. }

\section{Introduction}
Memristive devices are resistors exhibiting the effect of "memory" or hysteretic behavior in response to external field or driving, and have been shown to emulate well functions of biological synapses. An intriguing concept in the field is neuron-like synchrony and performance of circuits built of elementary memristive units. Significant progress in understanding
networks of memristive elements has been achieved recently in Ref.~\cite{Caravelli_complex_2017}. In addition, it has been strengthened that real (physical)
memristive systems may significantly differ from their model counterparts; see Ref.~\cite{Roldan,Wan_2022,Hwang}
for a~comprehensive review.

The generic model of a memristive system is \cite{Caravelli_memristors_2018}
\begin{subequations}\label{generic}
\begin{eqnarray}
\frac{d\mathbf{x}}{dt} &=& F(\mathbf{x},u)\\
y&=&H(\mathbf{x},u)\cdot u\,.
\end{eqnarray}
\end{subequations}
Here $\mathbf{x}$ is a vector of internal states.
The system has a memory, as the present-time values of $y$ depend, via $\mathbf{x}$, on past values of $u$.
If the pair $(y,u)$ is interpreted as current-voltage, $(I,V)$ we have a voltage-driven 
memristive system.

All the memristive systems discussed so far are assumed to be clean, or free of random perturbations.
In reality, all systems are perturbed and can be modelled in terms of stochastic processes, representing the inner fluctuations in the systems
or in  their outer environments. These fluctuations are termed noise and traditionally are described by Gaussian
White Noise (GWN) if they represent equilibrium fluctuations. Usually the effects of fluctuations 
are destructive as they blur or altogether destroy a coherent response of a system. However,
sometimes fluctuations, or noises, can act constructively. 
Stochastic Resonance (SR) is the best known phenomenon of this kind. 
In SR noise and a dynamical system act together 
to reinforce
a periodic signal \cite{Benzi,Nicolis,Hanggi}. The SR seems to be ubiquitous and has been claimed 
``an inherent property of rate-modulated series of events'' \cite{Bezrukov,dybiec_2009}. Several measures to quantify
SR have been proposed~\cite{Evstigneev_2005}; we are going to use the most popular one, namely, the Signal-To-Noise Ratio
(SNR) throughout this paper:
\begin{equation}\label{SNR}
\mathrm{SNR} = 10\log_{10}\frac{P_{\text{peak}}}{P_{\text{background}}}
\end{equation}
Here $P_{\text{peak}}$ stands for the power spectrum at the peak corresponding to the external
signal, and $P_{\text{background}}$ is the extrapolated background.
The interest in SR has 
now largely weaned, but it still remains an important feature of many noise-perturbed systems. And
surely enough, in memristive devices first a phenomenon of memory enhancement due to noise akin to SR
has been reported in Ref.~\cite{Stotland} and later a genuine SR in metal-dioxide memristors
has been investigated in Ref.~\cite{Spagnolo}. In addition, a model of stochastic resistance jumps
in memristive devices, not leading to SR, has been recently discussed in Ref.~\cite{slipko}.

Interestingly, similar effect have been observed in voltage-activated ion channels \cite{channel,ewa_2012,Bezrukov,rappaport}.
Ion channels, while not quite equivalent to memristive devices, share many of their features: the current passing
through an ion channel may depend on history and display hysteretic behavior, and gating dynamics governed by low and high conductance states have been shown to exhibit stochastic resonance~\cite{Bezrukov}. Hysteresis and memory effects are also important in such diverse contexts
as social systems~\cite{sznajd}, security devices~\cite{carboni} and many others that are too numerous to cite them here.

Notwithstanding previous works on constructive role of noises in nonlinear memristive systems~\cite{Roldan,Spagnolo,Stotland}, \egn{here we aim to discuss model systems where the quantitative analysis of memristic behavior is carried out within the framework of single- and multiple-well models of voltage driven switching in conductance. In particular, double-well stochastic models of that type mimick the Kramers theory of activated rate processes and have been successfully applied to analysis of hysteresis phenomenon in the conductance of voltage-sensitive ion-channels. \cite{ewa_2012,ewa_2019,rappaport,pustovoit,andersson}}
 These models belong also to the general \eqref{generic} class, with the internal parameter $\mathbf{x}$ being the dynamic conductance of the system, dubbed memductance in this context.
\egn{It will be demonstrated that incorporating fluctuations enhances current passing thorough the memristive device, thus showing a typical for the SR scenario, amplification of a (weak) signal by noise.}
\section{Dynamic memory in  voltage gated channels}
\egn{Ion channels are transmembrane, pore-forming proteins which regulate ionic currents through the cell membrane and undergo conformation deformations under environmental (temperature, electric field, pressure) changes. At the level of a single channel the gating process which changes permeability of ions in response to voltage change across the cell membrane incorporates also local conformational variations of the constituing proteins. Hysteresis, termed otherwise "mode shift" in ion channels is a phenomenon in which conductance loop arises in delayed response to voltage change, thus exhibiting a memory effect. Such hysteretic current-voltage characteristics has been detected in various biological channels \cite{KRUEGER2013,NOWAK1992,BENNEKOU2004} and the physiological significance of the phenomenon has been debated over the years \cite{andersson,Bezrukov}.
When the voltage varies sufficiently slowly, the protein constituent of the channel has enough time to adjust its conformation to the instantaneous value of the voltage. As a consequence, the ion current through the channel is independent of the prehistory and hence no hysteresis is observed. On the other hand, when the period of the voltage change is much shorter than the characteristic protein relaxation time, the protein molecule cannot follow fast variations of the voltage and adapts only to its average value. As a consequence, the current through the channel becomes again independent of the former history, and the hysteresis loop collapses to a single line. Altogether, the loop area first grows monotonically with the frequency of voltage change, reaches a maximum, and disappears as the frequency tends to infinity.
In models of a voltage-gated ion channel characterized by two states (open and closed) conductance $G(t)$ corresponds to the probability $P_O(t)$ of finding the channel in an open state \cite{andersson,Bezrukov,ewa_2019}. For $N$ uncorrelated channels the time dependent conductance obeys
\begin{equation}
G(t)=N[g_C+(g_O-g_C)P_O(t)]
    \end{equation}
where $g_i$ stands for the conductance of an individual channel in state $i$.
Time evolution of $G(t)$ is governed by combination of Eq.(3) and the rate equation
\begin{equation}
    \frac{dP_O(t)}{dt}= -k_O(t)P_O(t)+k_C(t)[1-P_O(t)]
\end{equation}
in which the kinetic rates $k_O(t)=k_O^*\exp[\alpha V(t)]$, $k_C(t)=k_C^*\exp[-\beta V(t)]$ are voltage-dependent and describe stochastic transitions between open and closed states. The memristive equation of the ion channel is then 
\begin{equation}
    I(t)= G(P_O(t), V(t), t) V(t)
\end{equation}
where, in general, the conductance may be voltage-dependent. In real channels, the voltage sensitivity is determined by a corresponding set of charged residues that react to changes in voltage potential, promoting conformational changes of the protein that, in turn,  generate discrete change of conductance \cite{Bezrukov, andersson,Ngo}}.

\section{An asymmetric double-well}
\egn{The free energy deﬁnes the work required to move from one region to another in phase space, such as e.g.  between  conformational states of a protein, or a channel operating between states of different conductance. In a recent study of the time dependent anion channel VDAC \cite{Ngo}, response to the transmembrane potential in the form of gating to low/high conductance states was analyzed by performing atomistic molecular dynamics studies. By choosing an appropriate collective "reaction coordinate" (be it e.g. conductance $G$), histograms of configurational states can be assembled and probability density of states $P(G)$ defined, see Fig.1. From the latter the free energy profiles can be estimated as $V_{\mathrm{eff}}(G)=-k_BT\ln{P(G)}$. }

\begin{figure}
\includegraphics[scale=0.75]{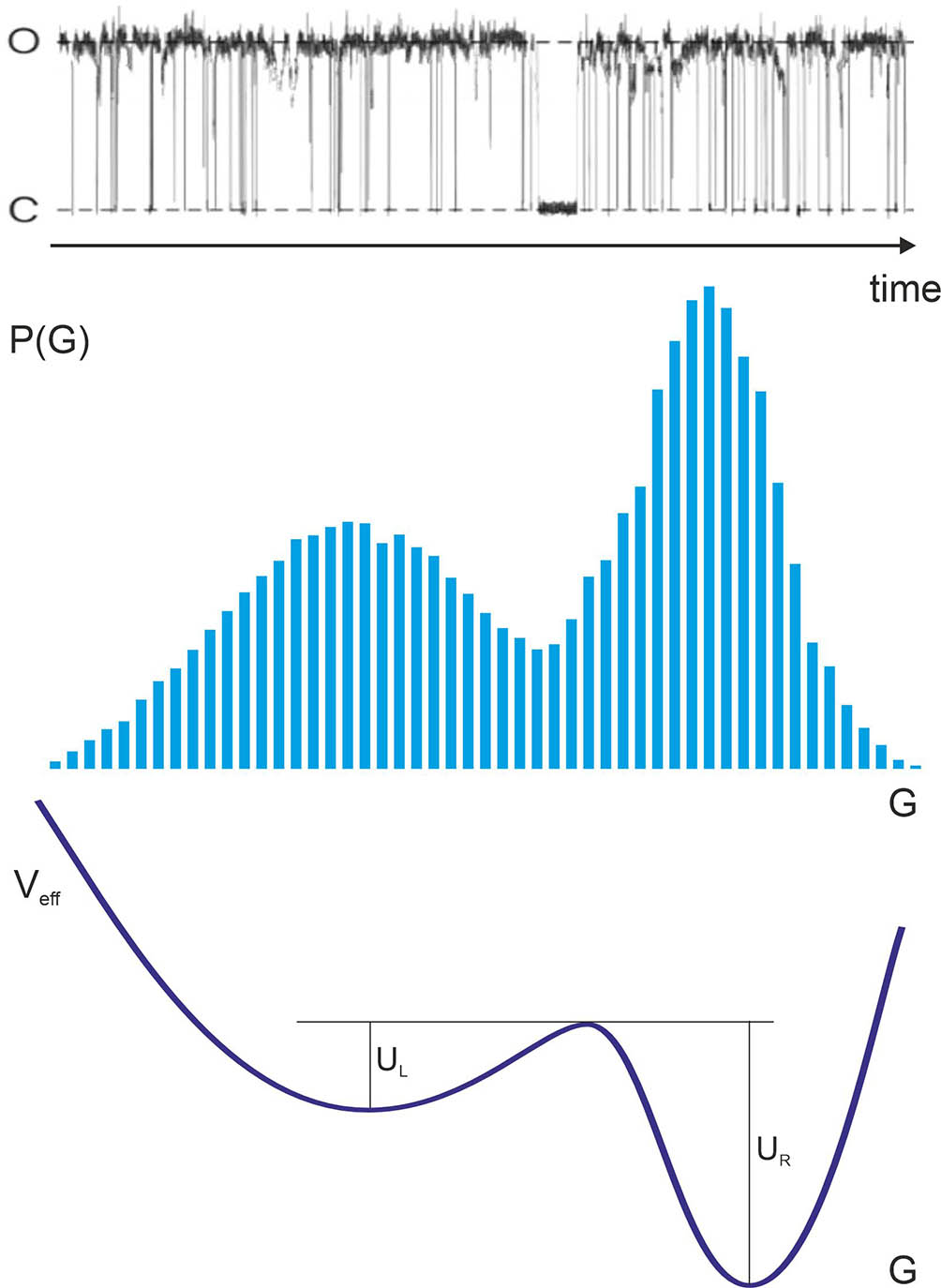}
\caption{Exemplary traces of a channel activity recorded at various conductance levels (C=closed state  O=open state) and histograms of conducting states derived from them (presented at  the right).
Bottom: an asymmetric effective potential  (free energy extracted from the stationary distribution) as
$-k_BTl\ln(P(G))$}
\end{figure}


Our first model refers to such an approach and assumes energy landscape of the channel gating in the form of the two well asymmetric (tilted) potential. Time changes in conductance follow then an overdamped motion in a 
 potential, possessing two minima, with 
respective depths $U_L,U_R$. 
The particular shape of the potential is of a lesser importance, 
but for the purpose of this research the following potential has been used:
\begin{equation}\label{potential}
V(G) = \frac{1}{4}(G-2)^4-\frac{1}{2}(G-2)^2 -\frac{1}{8}(G-2)\,,
\end{equation}
where $G$ is the memory-dependent conductance (memductance) of the system.
Such a configuration may be achieved by carefully doping a semiconductor. The equation of motion
is
\begin{equation}\label{rownanie}
\dot G = -\frac{dV}{dG} + V_1\cos{\omega t} +\sigma\,\xi(t)\,,
\end{equation}
and the associated current is
\begin{equation}\label{current}
I(t) = G(t)\cdot V_1\cos(\omega t)\,,
\end{equation}
where $\xi$ is a Gaussian White Noise and $\sigma$ represents its intentsity. The amplitude $V_1$ is
too small to drive the particle over the barrier. In this research, $V_1=0.2$ and $\omega=2\pi$.
In the absence of the external
voltage, $V_1=0$, the escape from a potential well forms a classic Kramers problem. Therefore, the ratio of
dwelling times
in the right and left potential wells is
\begin{equation}
\tau_R/\tau_L \sim \exp\left(\frac{U_R-U_L}{\sigma}\right).
\end{equation}

We assume that the system starts in states of low conductance (within the left potential well). Without the noise, the memductance would always remain there,
but thanks to the noise, 
it may cross the barrier and increase the current \eqref{current} 
transmitted by the system.
Once in the right well and provided the noise is \textit{not} very large,
the system has a tendency to stay there and perform
noisy oscillation around the deeper minimum.

\begin{figure*}
\includegraphics[scale=0.8]{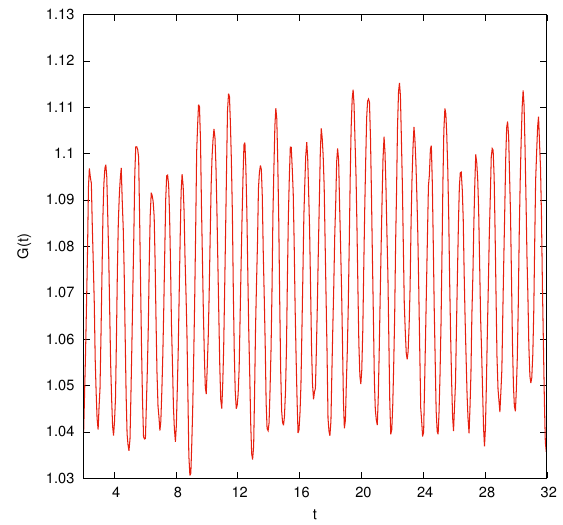}\includegraphics[scale=0.8]{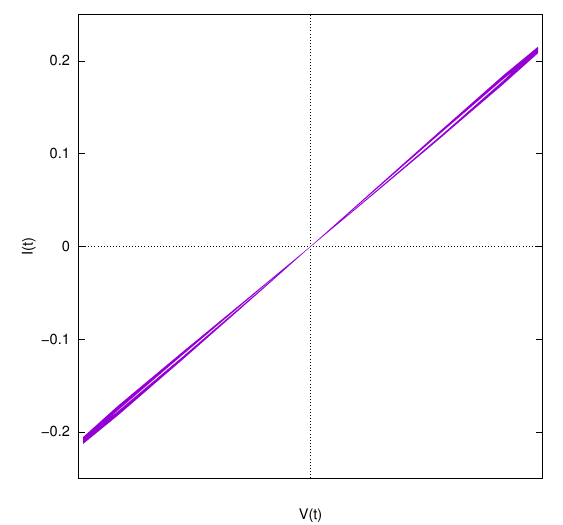}

\includegraphics[scale=0.8]{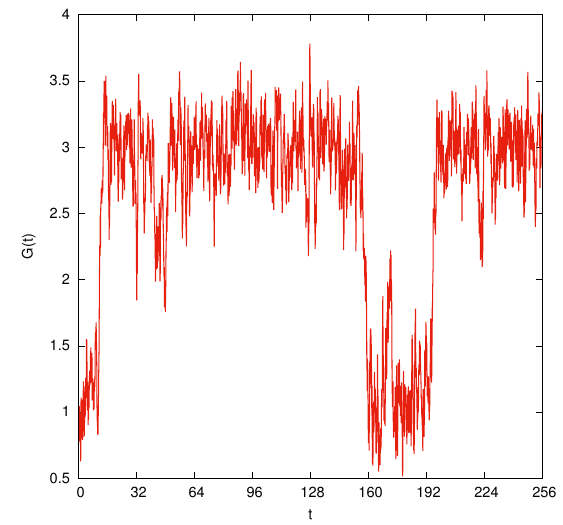}\includegraphics[scale=0.8]{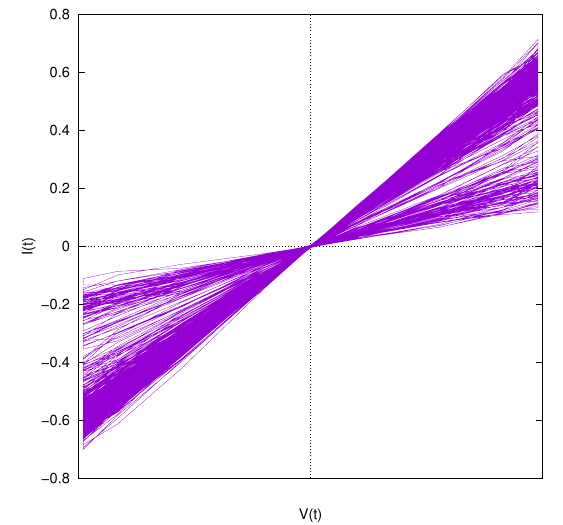}

\includegraphics[scale=0.8]{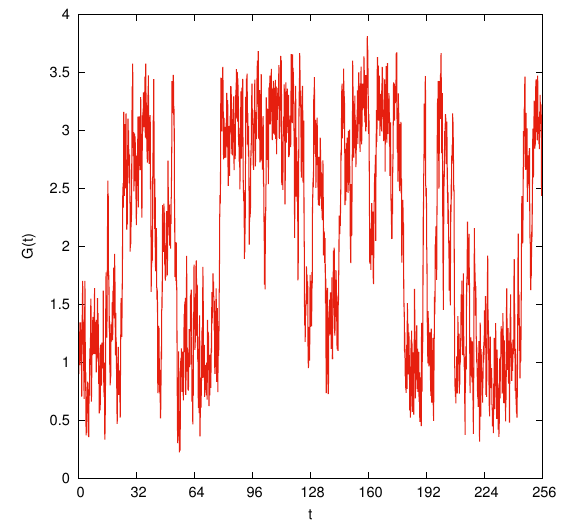}\includegraphics[scale=0.8]{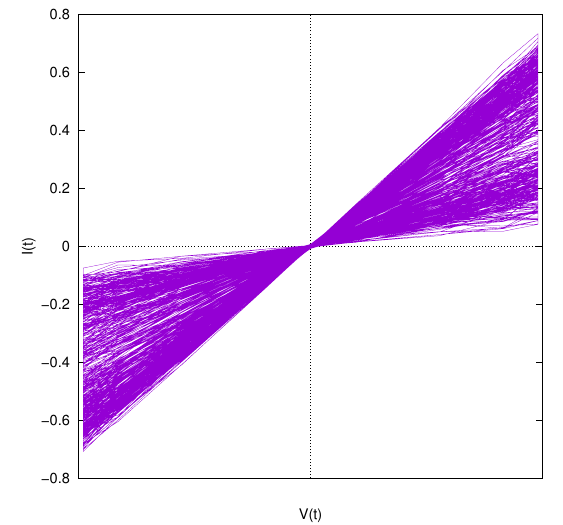}
\caption{Example trajectories (left column) and their corresponding hysteresis loops (right column)
for the system \eqref{rownanie}. Noise intensities are, top to bottom,
$\sigma=0.01$, $\sigma=0.5$, and $\sigma=0.75$.}
\label{panele}
\end{figure*}

\begin{figure}
\includegraphics[scale=0.9]{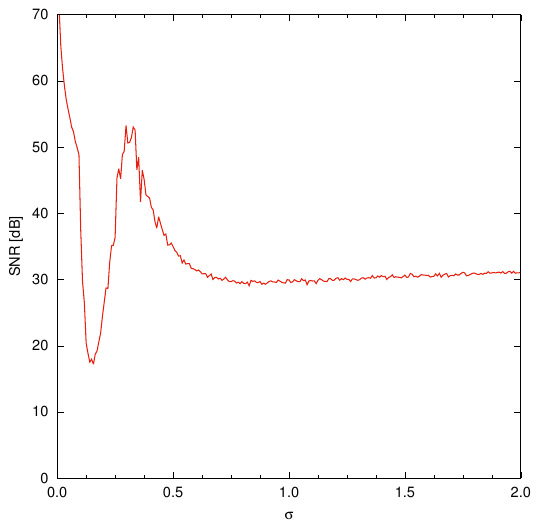}
\caption{The Signal-To-Noise Ratio in the system \eqref{rownanie} as a function of the GWN intensity.}
\label{peculiar}
\end{figure}

System \eqref{rownanie} has been solved numerically with the Euler-Maruyama method and a timestep $\Delta t=1/256$.
For the purpose of calculating SNR, trajectories have been averaged over 512 realizations for every single
value of~$\sigma$. 
For a very weak noise,
$\sigma=0.01$, the memductance displays oscillations in the left potential well. However,
consecutive minima and maxima are shifted due to the noise, and instead of a clear hysteresis loop, we
can see a collection of overlapping loops. For larger noise intensities, $\sigma=0.5$, and $\sigma=0.75$ 
the system crosses to the right potential well and spends most
of its time there. This means an increase of the memductance, but instead of a clear hysteresis loop,
we obtain a~combination of many overlapping, individual loops, bearing the impression of a~``dirty hysteresis'',
see Fig.~\ref{panele}.
For even larger intensities of the noise, the system \eqref{rownanie} ceases to see the fine structure of the
potential and the memductance performs a~random walk over all accessible range. This leads to
a~peculiar behavior of the SNR, see Fig.~\ref{peculiar}. For very small noises the system performs nearly perfect 
oscillations in the left potential well and the SNR is large. As the noise increases,
it gradually destroys the oscillations and the SNR decreases, finally reaching a~minimum.
After that, the constructive role of noise kicks in, helping the memductance to cross to the right
potential well in phase with the external voltage. The SNR reaches a~maximum
and a~Stochastic Resonance is observed.
However, for even larger values of the noise, the SNR displays
a~rather strange behavior: unlike in the regular SR, the SNR does not drop to zero, but reaches a~plateau 
that extends up to unphysically large
intensities of the noise. 

This last phenomenon requires, perhaps, some attention. The power spectrum of the current \eqref{current}
is related through the Wiener-Khinchin theorem to the autocorrelation
$\left\langle I(t)I(t+t^\prime)\right\rangle=\left\langle G(t)G(t+t^\prime)\right\rangle \cos\omega t\,\cos\omega(t+t^\prime)$.
If the noise is very large, the memductance $G(t)$ discontinues to see details of the potential and
is smeared over all accessible range. We may represent $G(t)$ in this regime as a random variable $\gamma(t)+\tilde G$,
where $\left\langle\gamma(t)\right\rangle=0$, the existence of $\left\langle\gamma(t)\gamma(t+t^\prime)\right\rangle$
is guaranteed by the fact that GWN is the driving process, and $\tilde G\not=0$ due to the asymmetry of the potential
with respect to sign reversal. Therefore,
\begin{equation}\label{korelacja}
\left\langle G(t)G(t+t^\prime)\right\rangle=\left\langle\gamma(t)\gamma(t+t^\prime)\right\rangle +{\tilde G}^2\,.
\end{equation}
The presence of ${\tilde G}^2\not=0$ is responsible for the plateau in the SNR.


\section{Harmonic well with correlated noises}

The other model discussed in this paper relates to the
Linear Stochastic Resonance (LSR) \cite{LSR}. LSR is a kind of SR where signal-enhancing effect arises from cooperation
between a linear transmitter and two GWNs acting on it, one multiplicative, or parametric, the other additive. 
Instead of \eqref{potential} we take
\begin{equation}\label{potencjalII}
V(G) = \frac{1}{2}aG^2 -V_0G\,.
\end{equation}
With the absence of any noises, this system leads to a time-delay behavior.
We now perturb the model by two GWNs: A~multiplicative, $\xi(t)$, and an additive, $\xi_a(t)$, noises.
The evolution equation for the memductance is
\begin{equation}\label{noise1}
\dot G(t) = -(a+p\xi(t))G + V_0 + q\xi_a(t) + V_1 \cos(\omega t+\phi)\,,
\end{equation}
where $\phi$ is a random initial phase of the signal.
The noises are correlated
\begin{equation}\label{correlation}
\left\langle\xi(t)\xi_a(t^\prime)\right\rangle=c\,\delta(t-t^\prime)\,,\quad -1\leqslant c\leqslant1\,.
\end{equation}
We may thus represent the additive noise $\xi_a(t)$ as a combination of two independent GWNs
\begin{equation}
\xi_a(t) = c\,\xi(t) + \sqrt{1-c^2}\,\eta(t)\,,
\end{equation}
with $\left\langle\xi(t)\eta(t^\prime)\right\rangle=0$, leading to
\begin{eqnarray}
\dot G &=& -(a+p\xi(t))G + V_0 + qc\xi(t) + q\sqrt{1-c^2}\,\eta(t)
\nonumber\\
&&{}+ V_1\cos(\omega t+\phi)\,.
\label{noise2}
\end{eqnarray}

The system \eqref{noise2} has been discussed in Ref.~\cite{LSR}.  The formal solution with $G(0)=0$
is
\begin{gather}
G(t)=\int\limits_0^t e^{-a(t-t^\prime)}\,
\exp\left[-p\int\limits_{t^\prime}^t\xi(t^{\prime\prime})dt^{\prime\prime}\right] \times{}
\nonumber\\
\left(V_0+qc\xi(t^\prime)+q\sqrt{1-c^2}\eta(t^\prime)+V_1\cos(\omega t^\prime+\phi)\right)dt^\prime\,.
\label{Gexact}
\end{gather}
This solution has a well-defined mean if
\begin{equation}\label{war1}
a-\frac{1}{2}p^2>0
\end{equation}
and a variance if a stronger condition
\begin{equation}\label{war2}
a-p^2>0
\end{equation}
holds. In this case, for $V_1=0$, the solution is
\begin{equation}\label{Ginfty}
\left\langle G(t)\right\rangle
\mathop{\longrightarrow}\limits_{t\to\infty}G_\infty=\frac{V_0-\frac{1}{2}cpq}{a-\frac{1}{2}p^2}
\end{equation}
and the variance asymptotically takes the form
\begin{gather}
\left\langle G^2(t) \right\rangle - \left\langle G(t)\right\rangle^2 
\mathop{\longrightarrow}\limits_{t\to\infty} 
\nonumber\\
D=
\frac{4V_0^2p^2-8aV_0cpq {+}(4a^2{-}4a(1{-}c^2)p^2{+}(1{-}c^2)p^4)q^2}{2(a{-}p^2)(p^2{-}2a)^2}\,.
\label{D}
\end{gather}
For $V_1\not=0$ we can calculate the correlation function
\begin{widetext}
\begin{equation}
\left\langle G(t)G(t+\tau)\right\rangle - \left\langle G(t)\right\rangle^2 
\mathop{\longrightarrow}\limits_{t\to\infty} 
\frac{V_1^2\cos(\omega\tau)}{2\left[(a-\frac{1}{2}p^2)^2+\omega^2\right]}
+
\left[\frac{V_1^2p^2}{4(a-p^2)\left[(a-\frac{1}{2}p^2)^2+\omega^2\right]} + D\right]e^{-(a-\frac{1}{2}p^2)\tau}
\label{Ctau}
\end{equation}
\end{widetext}
where now the braces additionally represent averaging over the initial phase of the signal.

\begin{figure}
\includegraphics[scale=0.75]{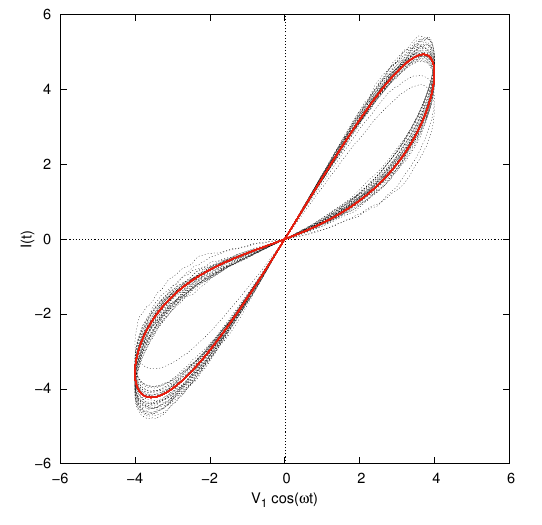}
\caption{\label{petlaa}Broken life --- the noisy hysteresis loop for $p=0.25$, $c=1$ and
with the condition~\eqref{resonance} satisfied. The other parameters are $V_0=1$, $a=1$, $V_1=4$, $\omega=2\pi$. 
The colored line 
shows the clean hysteresis loop corresponding to the same set of parameters.}
\end{figure}

The most interesting feature of the above solution is that if
\begin{equation}\label{resonance}
V_0p - acq=0
\end{equation}
for $c\not=0$ the variance $D$ reaches a minimum. This is so because with the condition
\eqref{resonance} satisfied, Eq.~\eqref{noise2} can be written as
\begin{eqnarray}
\dot G &=& -(a+p\xi(t))(G - V_0/a) + q\sqrt{1-c^2}\,\eta(t) 
\nonumber\\
&&{}+ V_1\cos(\omega t+\phi)\,.
\label{noise3}
\end{eqnarray}
As we can see, part of the additive noise translates only to a shift in the equilibrium solution
and $G_\infty=V_0/a$, as in the deterministic case.
Furthermore, if $c=\pm1$, the additive noise is eliminated altogether. For $D=0$ and without the
external signal, $V_1=0$, the device behaves as a noise-free resistor. When $V_1\not=0$,
$c=\pm1$ and the condition \eqref{resonance} satisfied, the solution is still noisy, but minimally so
for a given amplitude of the multiplicative noise,~$p$, see Fig.~\ref{petlaa}.

\begin{figure}
\includegraphics[scale=0.75]{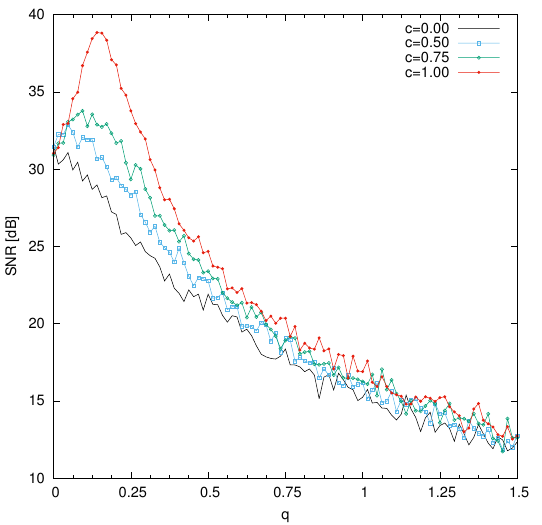} 
\caption{\label{SNRfigure}Numerically obtained Signal-To-Noise ratio as a function 
of the intensity of the additive noise,
$q$, for different values of the correlation coefficient, $c$
The multiplicative noise intensity $p=0.15$.
All other parameters as in Fig.~\ref{petlaa}.}
\end{figure}

\begin{figure}
    \centering
    \includegraphics[scale=0.75]{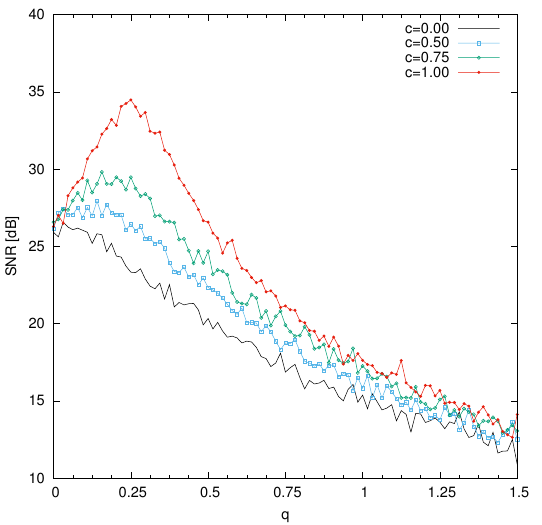}
    \caption{Same as Fig.~\ref{SNRfigure}, but with the multiplicative noise intensity $p=0.25$.}
    \label{SNRfigure25}
\end{figure}

\begin{figure}
\includegraphics[scale=0.75]{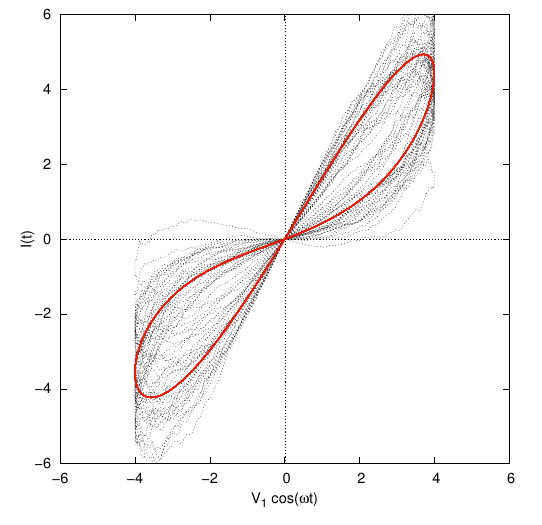}
\caption{\label{petlai}Same as in Fig.~\ref{petlaa}, but for $c=0$ and $p=q=0.25$.}
\end{figure}

Because $D$ enters the expression for the correlation function \eqref{Ctau}, minimizing $D$
corresponds to optimizing the current \eqref{current}
with respect to the external signal with amplitude of 
the multiplicative noise, $p$, fixed. We have solved Eq.~\eqref{noise2}
numerically with the Euler-Maryuama scheme with a timestep $\Delta t=1/256$. Numerical power spectra 
have been averaged over 128 realizations of the noises. Results for the model
\eqref{noise2} are presented in Fig.~\ref{SNRfigure}. For $c=1$, a clear stochastic resonance 
is visible. 
Stochastic resonance persists for all $c\not=0$, but for small values of $c$, SR is drowned by 
numerical fluctuations. For
uncorrelated additive and multiplicative noises, $c=0$, the stochastic resonance vanishes altogether
and the resulting hysteresis loop becomes much more irregular due to the maximization of the additive
noise. Trajectories representing negative memductances
are clearly visible, as shown in Fig.~\ref{petlai}.

Interestingly, for both amplitudes of the noises, $p$ and $q$, fixed, for $c\not=0$ 
the resonance condition
\eqref{resonance} can still be reached by changing $V_0$. This, however, means changing the 
shape of the noise-free hysteresis as well.

\section{Higher-order monostable wells}

\begin{figure}
\includegraphics[scale=0.75]{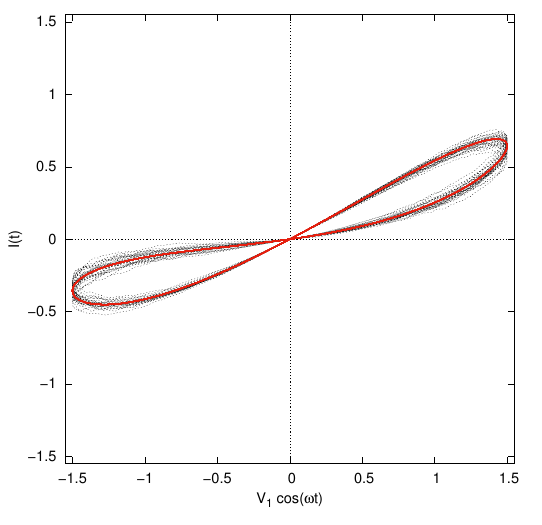}
\caption{ Broken life --- the noisy hysteresis loop for the quartic well,
\eqref{noise4} with $n=2$. $c=1$ and
the condition~\eqref{resonance} is satisfied.
Parameters are $V_0=1$, $a=1$, $p=0.15$, $V_1=1.5$, $\omega=2\pi$. The colored line 
shows the clean hysteresis loop corresponding to the same set of parameters. 
(Note a different scale as compared to Fig.~\ref{petlaa}.)}
\label{quarticpetlaa}
\end{figure}

\begin{figure}
\includegraphics[scale=0.75]{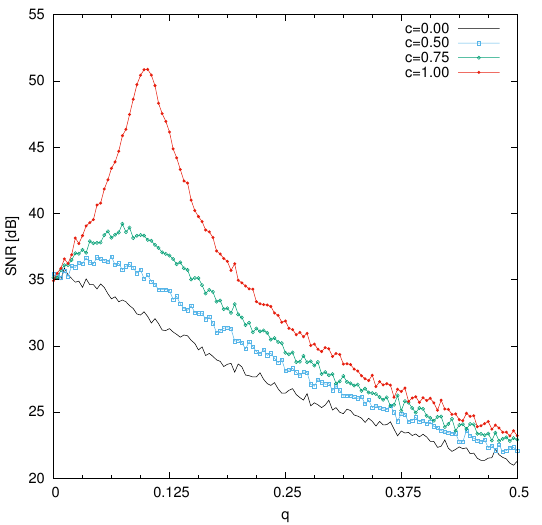}
\caption{\label{quarticSNRfigure}Numerically obtained Signal-To-Noise Ratio as a function 
of the intensity of the additive noise,
$q$, in the model \eqref{noise4}, for different values of the correlation coefficient, $c$. 
Other parameters as in Fig.~\ref{quarticpetlaa}.}
\end{figure}

Results for the LSR can be generalized to higher order monostable potential wells. Suppose
that instead of relaxing in a harmonic well, the particle whose position represents the
memductance relaxes in a potential $\frac{1}{2n}G^{2n}\,,n=2,3,\dots$. We now have
\begin{eqnarray}
\dot G &=& -(a + p\xi(t))G^{2n-1} + V_0 + qc\xi(t) + q\sqrt{1-c^2}\eta(t)
\nonumber\\
&&{}+ V_1\cos(\omega t+\phi)
\label{noise4}
\end{eqnarray}
and for $c\not=0$, with the condition \eqref{resonance} satisfied, 
\begin{eqnarray}
\dot G &=& -(a+p\xi(t))\left(G^{2n-1}-V_0/a\right) + q\sqrt{1-c^2}\eta(t) 
\nonumber\\
&&{}+ V_1\cos(\omega t+\phi)\,.
\end{eqnarray}
The intensity of the multiplicative noise, $p$, cannot be arbitrary large as the system would  become divergent. Unlike in the linear case, cf. Eq.~\eqref{war2} above, the maximal intensity can be assessed only numerically; it is always smaller than in the linear case.

Substituting $y=G-(V_0/a)^{1/(2n-1)}$ we get
\begin{equation}\label{noise5}
\dot y=-(a+p\xi(t))yW(y) + q\sqrt{1-c^2}\,\eta(t) + V_1\cos(\omega t+\phi)\,,
\end{equation}
where
\begin{equation}
W(y) = \sum\limits_{l=0}^{2(n-1)} 
\genfrac{(}{)}{0pt}{0}{2n-1}{l+1}
y^l 
(V_0/a)^{1-\frac{l+1}{2n-1}}
\end{equation}
is a polynomial of order $2(n-1)$. As we can see, correlations between the multiplicative
and additive noises again result in shifting of the equilibrium solution and reducing
the additive noise. If $c=\pm1$ and the resonance condition \eqref{resonance} is satisfied,
the additive noise is eliminated completely.

If there is no external signal, $V_1=0$, and the additive noise is eliminated, Eq.~\eqref{noise5}
is formally solved as
\begin{equation}\label{integral}
\int \frac{dy}{y\,W(y)} = -at -p\int\limits_0^t\xi(t^\prime)\,dt^\prime\,.
\end{equation}
For $n=2$, the integral on the left-hand of Eq.~\eqref{integral} can be carried out analytically,
but even then the resulting expression cannot be solved for $y$ explicitly. In general, 
Eq.~\eqref{noise4} can be solved only numerically. Because of the nonlinear character of
Eqns.~\eqref{noise4},\eqref{noise5}, a solution for $V_1\not=0$ cannot be represented as
a convolution of the free solution and the external forcing, as in the linear case discussed above.
However, as the integral in \eqref{integral} contains a logarithmic term, we expect that 
the condition \eqref{war2} needs to be satisfied for the variance of the solution to exist.
Numerical experiments confirm this intuition.

We have solved Eq.~\eqref{noise4} numerically for the quartic $(n=2)$ potential. The solutions
display a hysteresis loop, as expected. Numerical experiments show that the nonlinearity significantly reduces
the range of parameters for which negative memductances do not appear. Fig.~\ref{quarticpetlaa}
shows a hysteresis loop in a resonant case. Because with the amplitude of the multiplicative
noise equal $p=0.25$ as in Fig.~\ref{petlaa} leads to negative memductances, we have used
a smaller value of $p=0.15$ and the noisy hysteresis is less blurred.
Fig.~\ref{quarticSNRfigure} shows the SNR.
In a quatric well stochastic resonance is even stronger than in the LSR. SR is present for all
$c\not=0$, but for small values of $c$ it is hardly visible.

\section{Conclusions}
In this paper we have addressed the effect of noise on signal transmission in model memristive devices. On one hand side miniaturization of electronic processors allows for integration of many circuit elements per unit area and lowers the driving voltage, on the other hand, it naturally makes the systems vulnerable to thermal, $1/f$, shot and external, environmental noises. At the same time, based on theoretical approaches and experimental considerations, it has been documented that better endurance of signal transmission in natural and artificial systems can be achieved by understanding the source and making use of inherent temporal fluctuations in resistive devices. Contemporary methods used in design of artificial signal transferring systems try to mimick information processing mechanisms of living organisms and to emulate states of conductance of neuron synapses by analyzing stochastic response of neuron-like units and networks. Taken from that perspective, the Stochastic Resonance phenomenon may serve as an event optimizing performance of a system in the presence of noise, alike hearing or visual sensations have been shown to be amplified~\cite{Schilling,Simonotto,Hanggi,Dante_1997} by interference of weak signals and temporal fluctuations.

Our research shows that a conceptually very simple system of a particle relaxing in a potential well can model the memristive behavior under the influence of noise. We have discussed two kinds of models which, in addition to the memristive behavior, display a Stochastic Resonance. In the model involving a tilted double-well potential we observe a ``dirty hysteresis'' consisting of multiple overlapping hysteresis loops, reminiscent of hysteretic rounding observed in plastic or disordered materials.  In the model involving a monostable well subject to correlated multiplicative and additive noises -- a harmonic well that can be solved analytically and its generalisations to higher-order wells -- a blurred hysteresis is observed, much as in the case of hysteresis loops observed in voltage-activated ion channels~\cite{channel,ewa_2012,rappaport}. This highlights a previously unexpected connection between memristive systems and ion channels. The question remains how integration of such units in neuromorphic architecture will influence properties of the circuit and its performance in signal transduction.

\section*{Acknowledgments}

We would like to thank Benjamin Lindner for a most helpful discussion.
This work has been supported by
the Priority Research Area SciMat under the programme
Excellence Initiative -- Research University at the
Jagiellonian University in Krak\'ow.


\section*{References}
\bibliography{memrystory3}

\end{document}